# Identification for Accountability vs Privacy

19 January 2022 - Nick Pope (Security & Standards Associates)
Geoff Goodell (University College London)
Addendum added: 28 March 2022

## Abstract

This document considers the counteracting requirements of privacy and accountability applied to identity management. Based on the requirements of GDPR[1] applied to identity attributes, two forms of identity, with differing balances between privacy and accountability, are suggested, termed "publicly-recognised identity" and "domain-specific identity". These forms of identity can be further refined using "pseudonymisation" and as described in GDPR. This leads to the following forms of identity on the spectrum of accountability vs privacy:

- <u>Publicly-recognised identity,</u> which is a set of attributes that uniquely identify the person in the public domain. This provides high accountability but potentially exposes the person to privacy-related risks.
- <u>Unlinked domain-specific identity,</u> which is a set of attributes that only identify a person within the context of a specific domain and is not linked to a publicly-recognised identity. This maintains privacy but may have little or no accountability outside the domain.
- <u>Pseudonymised identity,</u> which is a domain-specific identity linked to a publicly-recognised identity within the domain. The set of attributes that publicly identify the person are managed within the domain separate from the domain-specific identity attributes associated with the pseudonym. This provides data protection and accountability outside the domain, contingent upon trust in the security (including business operations, and legal requirements) of data controllers and that they protect and do not misuse the identity attributes they manage.

An addendum to this paper considers examples illustrating the wider spectrum of identities with variations of the above 3 forms.

It is recommended that the privacy and accountability requirements, and hence the appropriate form of identity, are considered in designing an identification scheme and in the adoption of a scheme by data processing systems. Also, users should be aware of the implications of the form of identity requested by a system, so that they can decide whether this is acceptable.

## Privacy vs Accountability

One of the main aims of digital identities for natural persons, such as proposed under eIDAS[2], is to provide accountability of identified subjects for any actions on relying party systems. With digital identities, relying party systems can protect themselves against misuse, and in case of misuse occurring the natural person responsible can be traced and publicly held to account for their actions.

Privacy is also an important factor in the design of an identity management system. Under the GDPR (Regulation (EU) 2016/679) principle of data minimalization (Article 5 item 1c) states that "Personal data shall be ….. adequate, relevant and limited to what is necessary in relation to the purposes for which they are processed". Thus, identification related data (i.e. attributes) should only include any personal data which is necessary for the purpose to which the identity is used.

---

[1] Regulation (EU) 2016/679
[2] Regulation (EU) 910/2014 and proposed revision in COM(2021) 281 final

The requirements for accountability and privacy of identities should be carefully evaluated so that only minimal data required for the necessary accountability is released to a relying party. Depending on the form of identifier there is greater privacy or greater accountability as illustrated in the following diagram.

Two forms of identifier can be considered to represent the trade-off of accountability vs privacy:[3]

a) publicly-recognised identity which is a set of attributes that uniquely identify the person in the public domain. This could be for example a national identity, such as taxpayer identification number, driving licence number, travel document number, and so on, that is unique to the identified individual (e.g. is attributed to the individual's name as registered at birth, current registered address, date of birth, and so on). In cases where contact information, such as an email address, are publicly known to belong to an individual, this may also be used as a publicly-recognised identity. Such real-world identities provide direct accountability but include personal data.

b) domain-specific identity which is a set of attributes that only identify a person within the context of a specific domain. This could be for example a customer number or cookie used by a web site that allows anonymous connections and does not require its users to provide further identifying information. Provided that these attributes are not associated with information that could identify a particular user outside the domain, such domain-specific identities provide minimal accountability outside the specific domain but maximise the privacy of the data subject.

This trade-off between accountability and privacy is illustrated in the following graph:

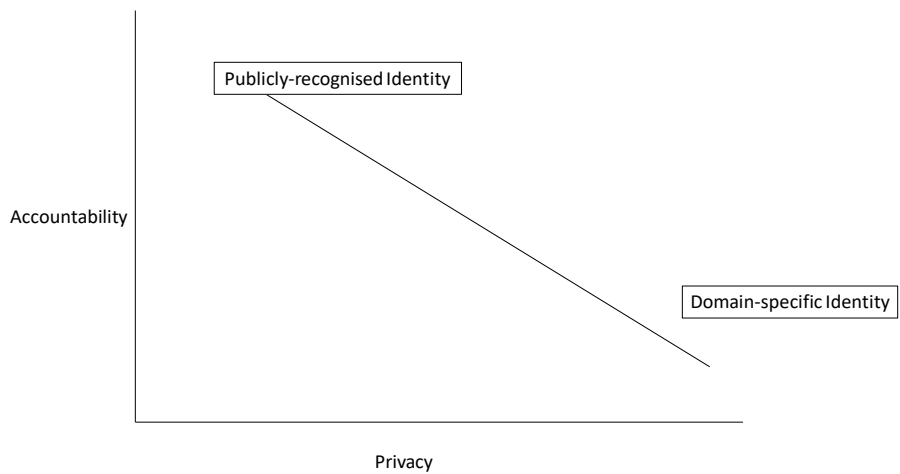

---

[3] Based on definition of identity given in ISO/IEC 24760-1.

- 2 -

## Pseudonymisation

According to GDPR (Article 4(5)): *'pseudonymisation' means the processing of personal data in such a manner that the personal data can no longer be attributed to a specific data subject without the use of additional information, provided that such additional information is kept separately and is subject to technical and organisational measures to ensure that the personal data are not attributed to an identified or identifiable natural person.*

Applying pseudonymisation to an identity, (i.e. using a pseudonym in place of a publicly-recognised identity), this can be taken to mean*: using an identifier [personal data] that can no longer be attributed to a specific subject without the use of additional information, provided that such additional information is kept separately and is subject to technical and organisational measures to ensure that the identifier [personal data] is not attributed to the [an] identified or identifiable natural person.*

A "pseudonymised identity" could be used by a data processing system for day-to-day transactions without being linked directly to personal attributes. Any personal attributes, as required by the for accountability, could be established when registering with the system and held separate from data relating to data-to-day transactions. Such personal attributes may be derived:

a) Through separate transactions, subject to appropriate measures for data protection, using a publicly-recognised identity to register with the system;
b) Through other means of identity-proofing such as physical or remote checking of identity documents.

## Recommended Forms of Identity

From the above discussion three basic forms of identity are recommended in this paper:

- Publicly-recognised identity, which is a set of attributes that uniquely identify the person in the public domain. This provides high accountability but potentially exposes the person to privacy-related risks.
- Unlinked domain-specific identity, which is a set of attributes that only identify a person within the context of a specific domain and is not linked to a publicly-recognised identity. This maintains privacy but may have little or no accountability outside the domain.
- Pseudonymised identity, which is a domain-specific identity linked to a publicly-recognised identity within the domain. The set of attributes that publicly identify the person are managed within the domain separate from the domain-specific identity attributes associated with the pseudonym. This provides data protection and accountability outside the domain, contingent upon trust in the security (including business operations, and legal requirements) of data controllers and that they protect and do not misuse the identity attributes they manage.

These three forms of identity can be illustrated on the accountability vs privacy spectrum as follows:

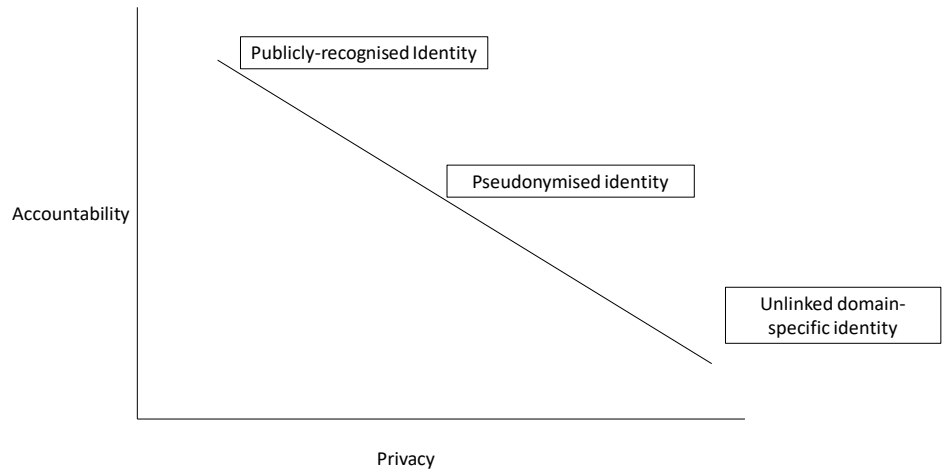

An addendum to this paper considers examples illustrating the wider spectrum of identities with variations of the above 3 forms.

It is recommended that the privacy and accountability requirements, and hence the appropriate form of identity, are considered in designing an identification scheme and the adoption of a scheme by data processing systems.  Also, users should be aware of the implications of the form of identity requested by a system, so that they can decide whether this is acceptable.

# Identification for Accountability vs Privacy
# Addendum- Illustrative Examples

22nd March 2022

This paper is an addendum to the earlier paper on Identification for accountability vs Privacy.  It illustrates the spectrum of identity with varying emphasis on Accountability vs Privacy by considering example scenarios across the spectrum.  These examples take into account the EU expert group Paper "EU Digital ID Architecture and Reference Framework – outline" but is not necessarily limited to identities supported by the EU Digital Identity regulatory framework.  This document makes assumptions on how identities might be presented under the framework which may not be reflected by the actual national implementations.

The following analysis is concerned with the impact on accountability and privacy that can occur through use of a particular form of identifier.  It does not address all privacy issues that may arise through use of an identity, such as information that may be derived through tracking the flow of information between network access points or information that may derived from the content of a transaction.

| Ref | Description | Presentation of identity to relying party | Accountability | Privacy |
|---|---|---|---|---|
| 1 | Full identity represented in a single certificate | [wallet IDr = 1234, Full name = Jean Schmidt, DoB = 1/2/2000 Country=Atlantis Auth = A1B2C3] SigAtl | The identity provides direct accountability of the entity. | Transaction includes data which publicly identifies the person and so affords no privacy. |
| 2 | Identity represented by wallet identity with set of verifiable statements linking attributes of Personal Identification Data with wallet. | [wallet ID = 1234, Auth = A1B2C3] SigAtl, [Subject = 1234, Full name = Jean Schmidt, Country = Atlantis ] SigAtl, | The identity provides direct accountability of the entity depending on additional attributes (date of birth) not revealed to the relying party but available in case of dispute. | Transaction includes some identification data but some sensitive data is not revealed to relying party. |
| 3 | Identity represented by wallet identity | [wallet ID = 1234, Auth = A1B2C3] SigAtl, | The identity provides direct accountability of the entity depending on additional attributes available in case of dispute. The transaction is also accountable within the domain through the re-use of the same wallet ID across multiple transactions. Attributes associated with a wallet ID can be made available to legal or regulatory authorities. | Although it reveals no additional identification data directly, the wallet ID is widely known by other relying parties as belonging to an identifiable entity. The wallet holder can be easily identified by other parties (authority, relying party, intermediary, or combination thereof) that has access to the attributes or personal information related to (or contained within) other transactions associated with this wallet ID. |

| Ref | Description | Presentation of identity to relying party | Accountability | Privacy |
| --- | --- | --- | --- | --- |
| 4a | When registering with a bank to open a new account the customer provides full details | [wallet ID = 1234, Auth = A1B2C3] SigAtl, [Subject = 1234, Full name = Jean Schmidt, Country = Atlantis ] SigAtl, [Subject = 1234, DoB = 01/02/2000 ] SigAtl [Subject = 1234, Address = 1 Pussy cat mews, Atlantis City] SigQTSP | The identity provides direct accountability in line with anti-money laundering requirements. | The registration includes data which publicly identifies the person and so affords no privacy. |
| 4b | The bank issues an customer identity number and establishes authentication data with the customer as identified in 4a. Subsequent payment transactions using a identity document issued by the bank. | [Customer ID = 5678, Auth = D1E2F3] SigBank | The transaction is accountable via the bank using the registration information provided to the bank. Attributes associated with a wallet ID can be made available by the bank to legal or regulatory authorities. | The payment transaction does not directly include identifiers recognised outside the bank. However, the bank can link the Customer ID to the account registration information, and so can aggregate information for a customer ID used across multiple transactions that reference that customer ID. |

| Ref | Description | Presentation of identity to relying party | Accountability | Privacy |
| --- | --- | --- | --- | --- |
| 5 | Gambling site issues an identity number not linked to any specific wallet or personal identification data.  The gambling site requires proof of age 21.<br>A QTSP issues an electronic attestation of age attributes for the gambler ID based on authentication of the subjects Personal Identification Data and authentication of the gambler ID issued by gambling website registered as a relying party. | [Gambler ID = 24589, Auth = F8E7D6] SigGam<br><br>[Gambler ID = 24589, Age = 22] SigQTSP | The gambling site cannot directly trace the transaction to the external identity of the user in a meaningful way.  However, the gambler can be traced via the QTSP that provided the age based on the gambler's external identity credentials.   Also, the transaction is accountable within the domain of the gambling site through the re-use of the same identity across multiple transactions. | Only age information is released to the gambling site.  However, the QTSP is aware of the association between the gambler ID and the person.  The gambling site can aggregate information from the customer throughout his or her use of the same gambler ID. |
| 6 | Website issues a "cookie", which is an identity number not linked to any specific wallet or personal identification data. | [Cookie = 24589, Auth = F8E7D6] SigWebsite | This transaction cannot be directly traced to the user in a meaningful way.  However, the transaction is accountable within the domain through the re-use of the same cookie across multiple transactions. | The Cookie in itself includes no identifying information but can be used to aggregate information about the user across all of his transactions. |
| 7 | Website uses a one-time identifier for each transaction, without attempting to identify the user across transactions.. | Transaction ID | The Transaction ID on its own provides no accountability. | The transaction ID in itself includes no identifying information. |

Notes regarding notation:

1. DoB = Date of birth
2. Auth = Data used to authenticate the identified entity (e.g. public key)
3. [ ….]SigAtl = Statement signed by Atlantis national authority
4. [,,,,,]SigQTSP = Statement signed by a QTSP
5. […..]SigBank = Statement signed using the bank's EU wallet registered as a relying party
6. […..]SigGam = Statement signed using the gambling site's EU wallet registered as a relying party
7. […..]SigWebsite = Statement signed using the website's EU wallet registered as a relying party